\documentclass[manuscript,nonacm]{acmart}

\AtBeginDocument{%
  }

\setcopyright{acmcopyright}
\copyrightyear{2024}
\acmYear{2024}
\acmDOI{XXXXXXX.XXXXXXX}

\begin{document}

\title{Hardware Trojan Detection Potential and Limits with the Quantum Diamond Microscope}

\author{Jacob N. Lenz}
\email{jlenz@mitre.org}
\orcid{0000-0003-3396-7536}
\affiliation{%
  \institution{Homeland Security Systems Engineering and Development Institute (HSSEDI), an FFRDC operated under contract by the MITRE Corporation for the Department of Homeland Security}
  \streetaddress{7525 Colshire Drive}
  \city{Mclean}
  \state{Virginia}
  \country{USA}
  \postcode{22102}}

\author{Scott K. Perryman}
\email{sperryman@mitre.org}
\orcid{0000-0002-3806-8348}
\affiliation{%
  \institution{Homeland Security Systems Engineering and Development Institute (HSSEDI), an FFRDC operated under contract by the MITRE Corporation for the Department of Homeland Security}
  \streetaddress{7525 Colshire Drive}
  \city{Mclean}
  \state{Virginia}
  \country{USA}
  \postcode{22102}
}

\author{Dmitro J. Martynowych}
\email{dmartynowych@mitre.org}
\orcid{0000-0001-6782-3675}
\affiliation{%
  \institution{Homeland Security Systems Engineering and Development Institute (HSSEDI), an FFRDC operated under contract by the MITRE Corporation for the Department of Homeland Security}
  \streetaddress{7525 Colshire Drive}
  \city{Mclean}
  \state{Virginia}
  \country{USA}
  \postcode{22102}
}

\author{David A. Hopper}
\email{dhopper@mitre.org}
\orcid{0000-0003-1965-690X}
\affiliation{%
  \institution{Homeland Security Systems Engineering and Development Institute (HSSEDI), an FFRDC operated under contract by the MITRE Corporation for the Department of Homeland Security}
  \streetaddress{7525 Colshire Drive}
  \city{Mclean}
  \state{Virginia}
  \country{USA}
  \postcode{22102}
}
  
\author{Sean M. Oliver}
\email{smoliver@mitre.org}
\orcid{0000-0003-3848-5632}
\affiliation{%
  \institution{Homeland Security Systems Engineering and Development Institute (HSSEDI), an FFRDC operated under contract by the MITRE Corporation for the Department of Homeland Security}
  \streetaddress{7525 Colshire Drive}
  \city{Mclean}
  \state{Virginia}
  \country{USA}
  \postcode{22102}
}

\renewcommand{\shortauthors}{Lenz et al.}

\begin{abstract}
The Quantum Diamond Microscope (QDM) is an instrument with a demonstrated capability to image electrical current in integrated circuits (ICs), which shows promise for detection of hardware Trojans. The anomalous current activity caused by hardware Trojans manifests through a magnetic field side channel that can be imaged with the QDM, potentially allowing for detection and localization of the effects of tampering. This paper seeks to identify the capabilities of the QDM for hardware Trojan detection through the analysis of previous QDM work as well as QDM physical limits and potential Trojan behaviors. QDM metrics of interest are identified, such as spatial resolution, sensitivity, time-to-result, and field-of-view. Rare event detection on an FPGA is demonstrated with the QDM. The concept of operations is identified for QDM utilization at different steps of IC development, noting necessary considerations and limiting factors for use at different development stages. Finally, the effects of hardware Trojans on IC current activity are estimated and compared to QDM sensitivities to project QDM detection potential for ICs of varying process sizes. 
\end{abstract}

\maketitle

\section{Introduction}

Ensuring the security of electronic devices is necessary due to their ubiquitous nature in applications spanning government, defense, finance, and other critical sectors\cite{TehranipoorWang2014}. At the heart of these devices are integrated circuits (ICs), which are developed by a segmented industry of vendors that conduct design, fabrication, post processing, packaging/assembly, and testing. These steps are often completed in different facilities by different vendors that then send the design/part on to others for further processing \cite{Chen2009}. Often these vendors and facilities are international, leading to further difficulty ensuring the security of chips transported outside their country of origin. The segmentation of the IC development process introduces a heightened vulnerability to IC tampering methods, such as the insertion of hardware Trojans. 

A hardware Trojan is any modification of the circuitry of an IC intended to alter its function. These alterations have been categorized as IC behavior modification, denial of service, or information leakage \cite{Chakraborty2009, Bhunia2014}. Small changes can lead to large security concerns, as they can cause the leakage of critical information like security keys or the degradation of critical processes like the generation of those secure keys \cite{Salmani2013, Shakya2017}. The example of compromised secure key generation highlights the possible adverse effects of a hardware Trojan and the security threat they pose \cite{Xiao2016}. 

Several methods for hardware Trojan detection have emerged to validate security during the IC development process. These methods are generally defined as either logic testing or side-channel analysis \cite{Bhunia2014}. Logic testing applies a set of directed test patterns intended to trigger the Trojan and propagate its effects to an IC output for detection. The vast number of possible triggers which could be used to activate a Trojan can make this method infeasible, necessitating a statistical approach like those detailed in \cite{ChakrabortyMERO2009,Jha2008}. Side-channel analysis measures incidental physical characteristics that change due to the activity of the IC, such as supply current, thermal activity, electromagnetic field (EM) emanations, and path delay, which are altered by the presence of a Trojan \cite{Bhunia2014}. These side-channel parameters can be affected by Trojan trigger monitoring, as the monitoring process leads to some change in the circuit's logical activity. This removes the necessity for ensuring activation of all required triggers to activate the Trojan, present in logic testing \cite{Ashok2022}. Side-channel detection methods of interest include measuring \textbf{supply current} by current monitors attached to power ports/pads \cite{Wang2008}, through the magnetic field side-channel it creates with devices like the Superconducting Quantum Interference Device (SQUID) magnetometer, or through backscatter analysis with an ensemble of probes \cite{Adibelli2020}, \textbf{thermal activity} monitoring with existing onboard thermal sensors or infrared (IR) thermography \cite{Forte2013, Nowroz2014, Salvi2021}, \textbf{EM emanations} measured in \cite{Ngo2015,Ngo2016} with an EM probe and oscilloscope, and \textbf{path delay} measured by on-board registers placed at the end of relevant paths \cite{Jin2008, Li2008}. Each of these techniques has certain advantages and disadvantages, but all suffer from the need for a Trojan-free sample (often referred to as a golden circuit) to distinguish anomalous from expected behavior in the presence of potential process variation and measurement noise \cite{Bhunia2014}. Research has begun to investigate the possibility of hardware Trojan detection without a golden circuit by utilizing machine learning methods to classify side-channel images of the device under test (DUT) \cite{Ashok2022}. 

The \textbf{Quantum Diamond Microscope (QDM)} has emerged as a viable solution for imaging magnetic field emanations from ICs for failure analysis \cite{Turner2020, Levine2020, Oliver2021, TurnerDiss2020, Basso2023, Levine2019} and shows promise for hardware Trojan detection \cite{Ashok2022}. The QDM is an imaging magnetometer that stands out due to its excellent spatial resolution, good sensitivity, wide field-of-view, and 3D vector readout capability. Magnetic field emanations from an IC can be detected and utilized to analyze the current in an IC. The QDM's spatial resolution lets an analyst spatially resolve areas of an IC. The wide field-of-view allows for imaging of large portions of the IC simultaneously, and 3D vector readout grants information that can remain hidden to a single vector tool like the SQUID magnetometer. In this manner, the QDM presents a tool with unique capabilities to add to the suite of microelectronics characterization techniques and a valid detection method for hardware Trojans in integrated circuits. 

This paper considers the use of the QDM for hardware Trojan detection and investigates the limits of the QDM for this application. We explore the relevant hardware Trojan categories in the context of a given threat model, from which we lay out a concept of operations as well as examples and explanations of the relevant imaging metrics for the QDM. QDM sensitivity and noise floor measurements are used to create projections of the minimum detectable current at varying measurement times. A demonstration of rare event detection was conducted to demonstrate the potential for detecting anomalous activity that occurs at short timescales (0.1 ms every 20 ms). This serves as a proxy for detection of Trojans that are active for brief windows of time. Finally, we estimated the current side-channels created by different IC process sizes and compared these to the minimum detectable current values to determine potential limits of Trojan detection with the QDM. 

\subsection{Threat Model}

\begin{figure}[t]
  \centering
  \includegraphics[width=\linewidth]{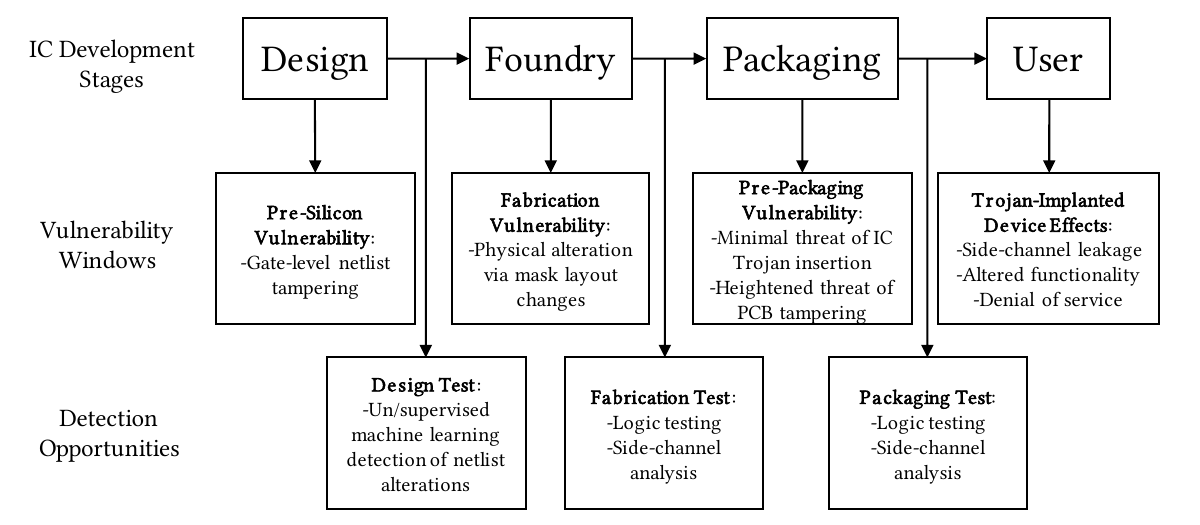}
  \caption{Integrated circuit (IC) Development Process: Avenues of vulnerability during IC development, where there is potential for the implantation of hardware Trojan and opportunities for Trojan detection. The IC development flow and potential windows of vulnerability are shown, as well as detection events where testing could break the flow of tampered devices to the user.}
  \Description{A flow chart depicting the IC development process and the potential threat entry points from each step, or avenues of vulnerability.}
  \label{fig1}
\end{figure}

The threat of tampering is present throughout the entire IC development process. In Figure \ref{fig1}, the top row shows the IC development flow, the middle row shows tampering opportunities during each stage of the fabrication process, and the bottom row shows detection breakpoints where testing could identify tampered devices. The risk of hardware Trojan implantation is compounded by the fact that these steps are often completed at different facilities and by different vendors, many times overseas from the eventual user. Potential tampering could be carried out by insiders at each stage of the process or by outsiders between the early stages of IC development. 

To ensure that a component is Trojan-free, it is important to test at each stage as detection becomes more difficult as the IC is brought closer to completion, as packaging added to devices can obscure side-channels and large sample subsets may be less available later in the development process. At the design stage, hardware Trojan layouts can be inserted into IC plans, or netlists, that are used to define the circuit layout and optimize IC parameters. During this pre-silicon development, supervised and unsupervised machine learning-dependent detection approaches have been identified that offer high efficiency and accuracy, but have failed to pair maximal accuracy with the ability to spatially determine Trojan activity \cite{Kok2019, Dong2020, Xie2017}. In the foundry where the IC is fabricated, the risk of hardware Trojan implantation emerges from physical alteration of the ICs themselves or the mask layouts used to fabricate the ICs as identified in Ref. \cite{Muehlberghuber2013}. The number of compromised devices affects detection efforts, as some subset of chips produced may be altered. The packaging stage often introduces a new facility and set of potential threats, with the opportunity for circuit modification before packaging is complete. Here the threat of IC modification is minimal, though, as the necessary tools are not readily available in most packaging facilities \cite{Harrison2021}. Testing at this stage is still important, especially if the design and fabrication stages are considered untrusted. Finally, a potential user could enact a final round of testing before placing a device into service. However, this may be the least efficient testing round as all devices are fully packaged and the user may have less inventory for testing than was available at the fabrication and packaging stages. 

The relevant threat model for QDM-based hardware Trojan detection in this work consists of the production of a subset of Trojan-implanted chips by a compromised IC development process before fabrication. In this scenario, the efficiency of the QDM must be assessed for detection at any point after the fabrication stage, including packaging and user testing stages. Thus, the threat model considered for this paper includes possible rounds of testing after fabrication, packaging, and user stages respectively. We assume access to the dies after production and the capability for destructive and non-destructive testing of any size subset of chips deemed necessary. This access changes depending on the stage considered for testing, and affects the type of testing protocol best suited to detection at a specific stage. 

\section{Hardware Trojans}

\subsection{Hardware Trojan Categories}

A hardware Trojan is defined as any modification of the circuitry of a microelectronic component to alter its intended function. Depending on the type and purpose of the component and the intentions of the tampering, hardware Trojan design can vary greatly in functionality and effect. Two key elements of a hardware Trojan's functionality are the Trojan's trigger and payload. The trigger activates the Trojan upon the fulfillment of some condition. The payload is any logic that is activated upon triggering to carry out the intended function of the Trojan. Trigger and payload logic can vary greatly in size, which is usually defined as some comparison to the intended (non-Trojan) logic of the circuit. This relationship depends heavily upon the intended logic within which the Trojan is implanted, and the trigger logic alone can be as small as 0.5\% of the total circuit logic of an AES cryptographic circuit design \cite{Ngo2016,Ashok2022}. 

A key Trojan trait is the Trojan's triggering characteristics. Trojans can be designed so that their payloads are always active or trigger upon the activation of user input, time-step, or random event \cite{Xiao2016}. Trojan triggers are especially important to consider when assessing a detection method. Timed and rare event triggers like those identified in Ref. \cite{Shakya2017} can affect the Trojan detection process as they may not be triggered during a particular detection method and thus may be missed. There are alterations to circuit activity due simply to the trigger monitoring, though, and thus these types of Trojans can be detected without triggering given a sufficiently sensitive detection method as demonstrated in Ref. \cite{Ashok2022}. Also of note, trigger logic for some hardware Trojans has been identified as larger than the activated Trojan logic, and thus detection is even more likely to be possible without triggering the payload \cite{Ngo2016}. 

The effect of a hardware Trojan, known as the Trojan’s payload, can take the form of altered functionality, information leakage through a side-channel, or denial-of-service/reduced functionality \cite{Shakya2017}. In these ways, the Trojan payload can change circuit function and determine the effect on system operation. A Trojan's payload may adversely affect detection efforts as the effects of activation may have a larger/smaller effect on the side channel used for detection. 

There is a large variety of potential Trojan trigger and payload combinations to accomplish specific effects on a compromised system. Known Trojan structures are collected in large part on Trust-Hub.org \cite{Shakya2017, Salmani2013}, with a taxonomy containing many example Trojans. Separating sample Trojan structures to better group them by triggers and payloads can help in understanding their eventual effects on compromised components. Hardware Trojans can be externally or internally activated, where external activation requires the attacker to access the device after its implementation and internal activation is triggered by some internal event or sequence. The threat models assessed for this work concern internally activated Trojans. These Trojans can be distinguished further as always-on or triggered (condition-based) \cite{Wang2008}. Potential combinations of internal activation states and payloads include (but are not limited to): 

\begin{itemize}

\item Always-On. Change in Functionality.
\item Always-On. Reduce in Functionality - Denial of Service.
\item Always-On. Side-Channel Leakage/Leakage Circuits.
\item Triggered (Time, Rare-Event, or User-Input Based). Change in Functionality.
\item Triggered (Time, Rare-Event, or User-Input Based). Reduce in Functionality - Denial of Service.
\item Triggered (Timer, Rare-Event, or User-Input Based). Side-Channel Leakage/Leakage Circuits.
 
\end{itemize}

\subsection{Hardware Trojan Detection}

The detection of hardware Trojans on ICs has been demonstrated via logic testing and side-channel analysis. Logic testing requires that a set of directed test patterns are carried out by the IC which may activate potential Trojans and propagate their effects to an output \cite{Chakraborty2009, Bhunia2014, ChakrabortyMERO2009, Banga2008, Waksman2013}. This method is ideal for Trojans with very few triggering conditions and robust against measurement noise and process variation but limited when considering Trojans with a large number of trigger conditions \cite{Bhunia2014}. This limits logic testing methods considerably, since Ref. \cite{Chakraborty2009} shows that even a Trojan constrained to four trigger nodes and one payload node in a circuit of 451 gates, there are $\sim10 ^{9}$ potential triggers.  Side-channel analysis on the other hand exploits physical parameters of IC activity like supply current or thermal radiation to reveal Trojan presence through abnormal circuit activity. The presence of a Trojan could result in the deviation of the current drawn on an idle circuit as well as the switching current profile of a circuit under test. These side-channel methods are more effective than logic testing when a Trojan has many triggers but are potentially constrained by process variation and measurement noise \cite{Bhunia2014}. 

IC side channels that give information on the activity of an IC include supply current \cite{Wang2008}, thermal radiation \cite{Forte2013, Nowroz2014}, EM field emanations \cite{Ngo2015}, and path delay \cite{Jin2008}. Measurement of supply current has been accomplished with current monitors attached to power ports/pads as in Ref. \cite{Wang2008}, by measuring the magnetic field side-channel due to current \cite{Ashok2022}, and via EM field backscattering which receives a scattered signal from a set of probes \cite{Adibelli2020}. Thermal activity has been measured using existing onboard thermal sensors in concert with predictive thermal models as in Refs. \cite{Forte2013, Nowroz2014} or via IR thermography as in Ref \cite{Salvi2021}. EM field side-channels can be measured with a simple EM probe and oscilloscope as in Refs. \cite{Ngo2015,Ngo2016} and path delay side channels can be measured using onboard registers placed at the end of relevant paths \cite{Jin2008, Li2008}. The activity indications resulting from each side-channel analysis technique can reveal the presence of a hardware Trojan through comparison to a golden circuit. This golden circuit requirement can be simple for an organization operating at the fabrication stage and more difficult for an end user that wants to verify a small subset of ICs they have received as they may not have ready access to a golden circuit. 

An important aspect of these methods of side-channel analysis is how well the side channel can be read non-invasively. Testing is most likely easier at the fabrication stage before packaging and encapsulation. For an end-user with a smaller batch of chips to test, decapsulating and destroying a portion of them for testing may be impractical. Power draw and path delay side channels may be more optimal for testing after packaging, as they are more easily detected non-invasively. The magnetic field side-channel created by supply current is also notable due to its through-package potential, as magnetic fields generally pass through IC and packaging materials \cite{Li2021}. The magnetic field created by the anomalous supply current caused by a Trojan in an intact or decapsulated IC could be detected by a sufficiently sensitive magnetometer. A reliable method for non-invasive side-channel analysis would allow for Trojan detection throughout the IC development process.

\section{Quantum Diamond Microscope}

\begin{figure}[t]
  \centering
  \includegraphics[width=\linewidth]{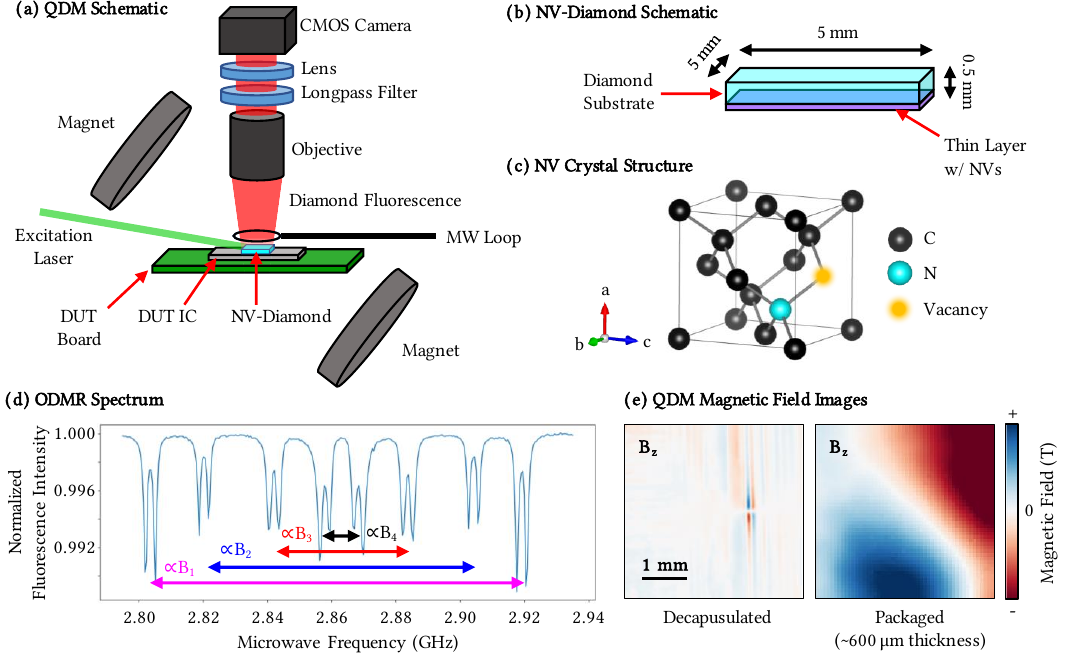}
  \caption{(a) Quantum Diamond Microscope (QDM) schematic. The instrument is built around a diamond chip embedded with a thin layer nitrogen-vacancy (NV) center defects. The NV-diamond can be placed directly on the device under test (DUT) to maximize spatial resolution in magnetic field images. (b) NV-diamond schematic. These diamonds are typically $\sim$5 mm x 5 mm x 0.5 mm and consist of a pure diamond substrate and a thin layer containing an ensemble of NV centers. (c) NV center atomic structure. An NV center consists of a substitutional nitrogen atom (cyan sphere) and neighboring lattice vacancy (orange sphere) in the diamond carbon lattice (black spheres). Based upon the potential location of the vacancy with respect to the nitrogen atom, there are four possible orientations of the NV axis in the diamond lattice. (d) An example of a typical QDM optically-detected magnetic resonance (ODMR) spectrum, which is collected per pixel with the instrument’s camera. Fluorescence is measured as a function of applied microwave (MW) frequency. The frequency separation of the four resonance feature pairs (dips in the spectrum labeled with magenta, blue, red, and black arrows) is proportional to a magnetic field’s projection (B$_{1}$-B$_{4}$) onto the four possible orientations of NV axes in the diamond lattice. (e) QDM B$_{z}$ magnetic field images of ring oscillators active on a field programmable gate array (FPGA) showing that the instrument can detect and localize functional activity on both a decapsulated and intact chip.}
  \Description{DESCRIBE IMAGES}
  \label{fig2}
\end{figure}

The QDM is a magnetometer capable of acquiring images of magnetic field emanations from microelectronics to localize functional activity (see the schematic in Figure \ref{fig2}(a)) \cite{Turner2020, Oliver2021, Ashok2022}. In the case of ICs and other microelectronics components, current flowing through a device produces magnetic fields that can be detected using a magnetometer. The QDM has a demonstrated wide field-of-view, high spatial resolution, and allows for vector magnetic field imaging under ambient conditions (no cryogenics needed) \cite{Oliver2021, Turner2020, TurnerDiss2020}. 

The QDM is built around a synthetic diamond chip embedded with an ensemble of defects known as nitrogen-vacancy (NV) centers that are distributed in a thin layer across one face of the diamond (typically 1-100 microns thick). A schematic of the QDM’s NV-diamond chip is shown in Figure \ref{fig2}(b) and the crystal structure of a diamond NV center is shown in Figure \ref{fig2}(c). See Ref. \cite{Edmonds2021} for details on growth of NV-diamond for sensing. As shown in Figure \ref{fig2}(c), a single NV center in diamond consists of a substitutional nitrogen atom (cyan sphere) and a neighboring lattice vacancy (orange sphere) embedded in the diamond’s carbon lattice (black spheres). Due to the structure of the diamond crystal lattice, there exist four possible orientations of the NV axis (line connecting the nitrogen atom to the vacancy) with respect to the nitrogen atom. The NV concentration is typically on the order of 1-15 parts per million \cite{Edmonds2021} and it is assumed that there is a homogeneous distribution of NVs with their axes in the four possible orientations across the NV epilayer.

Each of the diamond's NV centers act as an atomic-scale magnetometer that emits a magnetic field-dependent fluorescence which can be read out optically. Magnetic fields emanating from a sample interact with the NV centers, changing their electron spin states, which changes their optical emission. The fluorescence is spatially resolved with a camera, analyzed, and converted into maps of the vector components of a magnetic field. This is in contrast to other magnetometers like SQUID magnetometers that are only sensitive to one component of the magnetic field.

High spatial resolution magnetic field imaging with the QDM requires that the NV centers be as close to the magnetic field sources as possible. The magnetic field intensity decreases with standoff distance $r$ as $1/r$ or $1/r^{2}$ (depending on sensor-sample standoff distance), and large standoff distances act as a lowpass filter resulting in loss of high spatial frequency information \cite{Levine2019, Lima2009}. Since the QDM does not require cryogenics (like a SQUID magnetometer) and the diamond is chemically inert, small standoff distance ($\sim$1-100 µm) can be achieved by making the diamond touch the DUT if necessary. Spatial resolution can also be improved by minimizing the thickness of the diamond's NV-containing epilayer, which reduces the average distance from the NV centers to the DUT. 

QDM measurements collect a continuous wave optically-detected magnetic resonance (CW-ODMR) spectrum simultaneously at each pixel of the instrument’s camera. A typical CW-ODMR spectrum is shown in Figure \ref{fig2}(d). In this measurement protocol, NV center fluorescence (red emission, centered at 637 nm) is induced through excitation with a continuous wave green laser (in the absorption band of 500-660 nm, typically 532 nm is used) and is monitored as a function of applied microwave (MW) field frequency. Laser excitation can be carried out at a shallow angle to maximize internal reflection in the diamond, and therefore maximize NV center excitation. Fluorescence is collected with an objective, passed through a longpass filter and lens, and spatially resolved with a CMOS camera (see the optical path in Figure \ref{fig2}(a)). A MW loop near the diamond applies a field with varying frequencies to the NV centers. As shown by the magenta, blue, red, and black arrows in Figure \ref{fig2}(d), 4 pairs of resonance features can be measured in an ODMR spectrum (there are two peaks per resonance feature due to interactions of the NV center electrons with the nitrogen nucleus). The frequency separation between the low and high frequency for each pair (marked with an arrow in Figure \ref{fig2}(d)) is proportional to the magnetic field strength B$_{i}(i=1,2,3,4)$ projected onto the four NV axis orientations. The vector components of a magnetic field in the laboratory reference frame (B$_{x}$, B$_{y}$, B$_{z}$) are then calculated through a series of linear combinations of the magnetic field in the diamond reference frame B$_{1}$, B$_{2}$, B$_{3}$, and B$_{4}$. For more information on NV physics, see \hyperlink{appendix.A}{A Appendix: NV Physics}.) In this conventional CW-ODMR sensing protocol, each of the 8 resonance features are sampled as the MW frequencies are swept across to create images of the magnetic field in 3 dimensions. There is a technique for reducing the number of MW frequencies needed to image activity, termed lock-in sensing, where one or two of the resonant frequencies can be selected and the relative magnetic field can be determined by testing a small number (1-16) of frequencies around the chosen resonance features. This can be advantageous for applications where fast imaging is preferable and it is not necessary to image all 3 vector components.

In cases where more exact spatial resolution of the current path in a DUT is required, it may be beneficial to invert magnetic field images to obtain maps of the current distribution. This requires solving the magnetic inverse problem, where the Biot-Savart Law for magnetic field \textbf{B} as a function of current density \textbf{J} is inverted to solve for \textbf{J}. While a unique solution exists for two-dimensional current distributions, the problem is ill-posed and therefore has a non-unique solution in three dimensions. Several techniques have been developed for constraining and solving the magnetic inverse problem such as the Tikhonov-projection scheme \cite{Kress2002}, Fourier filter formalism \cite{Roth1989,Tan1990}, estimation theory \cite{Hamalainen1994}, probabilistic multi-source reconstructions \cite{Greenblatt1993}, least square fitting \cite{Hamalainen1993,Gonnelli1987}, Bayesian methods \cite{Auranen2005}, genetic algorithms \cite{Netter2001} and direct mapping and fitting in low dimensionality systems \cite{Johansen1996,Zuber2018,Jooss1998}. Demonstrations of failure analysis via magnetic side-channels have avoided the inverse problem by placing reasonable constraints on the current paths available based on relevant geometries and previous knowledge of the DUT \cite{Gaudestad2014,Kor2011,Felt2007}. In some cases, such as hardware Trojan detection and qualitative failure analysis of ICs, a precise reconstruction of current paths in the DUT may not be necessary. The user may only care about localizing activity on a chip, which can be done by identifying areas of increased magnetic field activity. This typically requires comparing a magnetic field measurement of a chip in its ON versus OFF state (ON is current flowing through the desired circuit, OFF is the absence of that current), comparing images of a test chip and a golden circuit, or comparing the measurements from a DUT occupying multiple functional states.

\subsection{QDM Imaging Metrics}

The QDM is often characterized by its spatial resolution, sensitivity, field-of-view, and time-to-result, as these metrics demonstrate the advantages of the QDM for imaging IC activity. Below we have described these metrics and provided an example of spatial resolution, a demonstration of small timescale event detection, and projections of QDM sensitivity at different time-to-results. 

\subsubsection{Field-of-View and Spatial Resolution}

The QDM’s field-of-view is limited to the lateral dimensions the diamond, which are typically on the order of 5x5 mm$^{2}$. Diamonds have been grown up to $\sim$10x10 mm$^{2}$ \cite{Balmer2009}, but challenges with large area growth and limited commercial demand for these larger diamonds makes the process expensive \cite{Achard2020}. Growth requires a diamond seed substrate, which limits the final diamond area since substrates are typically on the order of a few tens of mm$^{2}$. Additionally, diamond growth in the lateral dimension often results in twinning and defect nucleation from corners, which induces cracks and stress \cite{Tallaire2013}.

\begin{figure}[t]
  \centering
  \includegraphics[width=\linewidth]{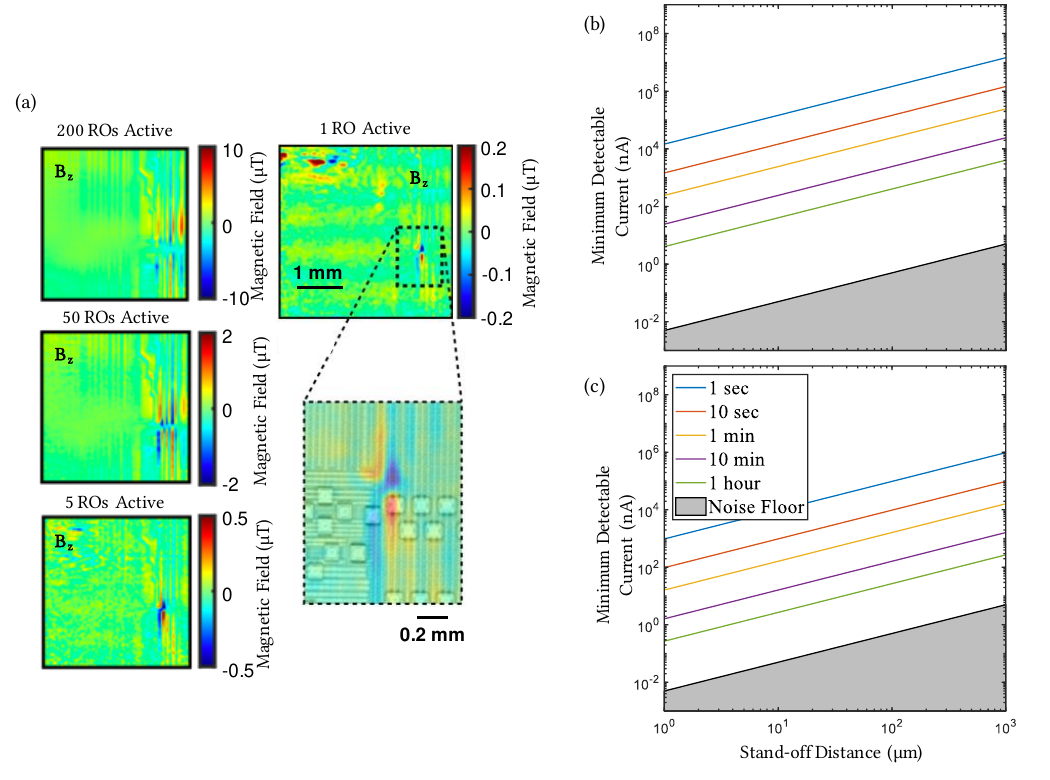}
  \caption{(a) Quantum Diamond Microscope (QDM) magnetic field images from Ref. \cite{Turner2020} demonstrating a spatial resolution of $\sim$10 µm. These images were taken of an FPGA (Artix-7, 28 nm process size) with varying numbers of active ring oscillators (ROs), and highlight the QDM's ability to detect and localize signals from minimal circuit activity as each RO utilizes six transistors. (b)-(c) Plots of projected values for minimum detectable current (nA) at varying total measurement times, based on the noise floor and magnetic field sensitivity possible with current QDM setups. (b) depicts the minimum detectable current at each total measurement time for a full optically detected magnetic resonance (ODMR) spectrum, while (c) depicts the projected minimum detectable currents for the application of a lock-in method that improves sensitivity by reducing single measurement time. Both (b) and (c) depict an environmental magnetic noise floor (greyed out area) that acts as a lower bound to current detection without the use of electromagnetic shielding. Plots in (b) and (c) share the same x axis.}
  \Description{DESCRIBE IMAGES.}
  \label{plot1}
\end{figure}

Spatial resolution of the QDM depends on standoff distance \cite{TurnerDiss2020,Levine2019}. Large standoff distance acts as a lowpass filter, resulting in a loss of high spatial frequency information and a blurring of the magnetic fields \cite{Lima2009}. In addition to the distance from the diamond to the DUT, the NV layer thickness also affects standoff distance as the average separation of the individual NV centers from the sample is approximately one-half the thickness of the NV layer. Typically, a naive assumption is made for magnetic field imaging that the spatial resolution $s$ is roughly equal to the standoff distance $z$, although other magnetic field imaging techniques like the SQUID have shown to have a spatial resolution of $s \sim z/5$ \cite{Wellstood2003}. Lastly, a lower bound to the QDM spatial resolution will always be set by the optical diffraction limit $\lambda/(2*NA)$, where $\lambda$ is the wavelength of light detected (NV center fluorescence is $\sim$637 nm) and NA is the numerical aperture of the QDM’s objective. This gives an estimated lower bound of 2.45 µm for NA=0.13, and can improve to 0.228 µm with alternative objectives such as those with oil immersion lenses (NA=1.4). As highlighted in Figure \ref{plot1}(a), recent demonstrations have shown that QDM magnetic field images can obtain a spatial resolution of $\sim$10 µm, allowing for the resolution of current activity on an Artix-7 which was built using 28 nm process size technology\cite{Turner2020}.

\subsubsection{Time-to-Result and Current Sensitivity}

Time-to-result for obtaining meaningful QDM magnetic field images is an important metric of performance and different applications have different requirements. The physics of NV centers puts a lower bound on the temporal resolution of the QDM. Detection of a magnetic field is carried out by optical readout of NV centers’ electron spins as they are excited and relax between a ground and excited electronic state and an alternative spin-dependent decay pathway (known as a singlet state). The limiting factor is the $\sim$140-200 ns lifetime of the singlet state, as opposed to the shorter 13 ns lifetime of the excited state, that sets a lower bound of $\sim$5 MHz on the QDM sampling rate \cite{TurnerDiss2020, Barry2020, Robledo2011, Gupta2016, Acosta2010}. Instrumentation and measurement limitations make it challenging for the QDM to reach this theoretical limit on temporal resolution.  

Figure \ref{plot1}(b) shows projected QDM minimum detectable current for different time-to-results when utilizing current CW-ODMR capabilities, and \ref{plot1}(c) shows the minimum detectable current when utilizing a digital lock-in method which reduces the number of frequencies tested during each run. Three distinct measures of time can be identified regarding measurements with the QDM: exposure time, single measurement time, and total measurement time. The exposure time is the amount of time the camera captures during each measurement (usually between 3-20 ms) and does not include the overhead time (such as the time necessary to change the state of the IC between runs). The single measurement time is the amount of time required to measure the ODMR spectra at each pixel (including overhead), determined by the number of MW frequencies tested in the range around each resonant frequency. The total measurement time is determined by the number of single measurements averaged which can be scaled up to decrease the effects of systemic noise, bringing the sensitivity closer to the environmental noise floor shown as the greyed out area in Figure \ref{plot1}(b) and (c). The projections in Figure \ref{plot1}(b) and (c) utilized the magnetic field sensitivity and noise floor of a current QDM setup, with a voxel size of 3 µm x 3 µm x 10 µm leading to a volume normalized sensitivity of $\sim$6.7 µT µm$^{3/2}$Hz$^{-1/2}$. Reducing the single measurement time improves the sensitivity, as sensitivity scales with $\sqrt{\tau}$ where $\tau$ is the single measurement time \cite{Hopper2018}. The lock-in improvements in minimum detectable current shown in Figure \ref{plot1}(c) were estimated based on the improvement in magnetic field sensitivity due to a decrease in necessary single measurement time, measuring only a few resonant frequencies rather than a full sweep and allowing for more averaging over the same period of time, which agrees with other QDM implementations \cite{TurnerDiss2020}. 

Different applications have different requirements, which dictates QDM measurement modality and therefore time-to-result. Different instances are described below:

\begin{itemize}

\item \textbf{Careful quantitative mapping of all vector components of a magnetic field:} The primary advantage of the QDM is the wide-field, simultaneous imaging of the B$_{x}$, B$_{y}$, and B$_{z}$ vector magnetic field components, which allows an analyst to measure both in-plane and out-of-plane current distributions that may arise in certain devices, like 3D ICs, and allows for detection of current sources that may have canceling or masked B$_{z}$ fields. For these measurements typically about 50-60 MW frequencies centered around the 8 resonances in Figure \ref{fig2}(d) are applied to sufficiently sample the curves for fitting and extraction of their center frequencies. These measurements take the longest, with a time-to-result of minutes to tens of minutes for an image with a signal-to-noise ratio (SNR) of 5-10, which is typically necessary for current analysis. (See  \hyperlink{appendix.B}{B Appendix: Signal-to-Noise Ratio} for further information on achievable SNR.)

\item \textbf{Careful quantitative mapping of the magnetic field without vector information:} In some cases, activity mapping in an IC may require carefully measured, single vector magnetic field information. Here, MW frequencies only need to be swept across a single pair of resonance features (one of the resonance pairs highlighted by the magenta, blue, red, and black arrows in Figure \ref{fig2}(d)) instead of all 4 pairs, thereby reducing the measurement time by a factor of 4. Here, the magnetic field projection is measured for only one orientation of the 4 NV axis orientations. Due to the 109.5$^{\circ}$ tetrahedral bond angle of the diamond lattice, single axis measurements are sensitive to not only the out-of-plane component of the magnetic field but in-plane components as well.

\item \textbf{Qualitative, fast mapping of IC activity:} Often, there is little need for measurements of exact magnetic field amplitudes and instead an analyst only cares about localizing activity in an IC. Here, sparse sampling of MW field frequencies is sufficient to detect changes in magnetic field and allows for a significant reduction in measurement time. Instead of carefully tracing out the resonance features in the QDM’s ODMR spectra through application of hundreds of MW frequencies, only a few MW frequencies need to be sampled per NV resonance. For example, two MWs can be applied to a single NV center resonance  – one on the higher frequency side and one on the lower frequency side of the resonance selected to be at the location of maximum slope. The diamond fluorescence is spatially monitored at the two MW fields chosen. As magnetic fields from the DUT interact with the NV centers, the resonances shift in frequency, which changes the relative fluorescence at the two MW field frequencies. Images of the IC powered ON and OFF are saved, and the fluorescence changes are digitally subtracted, or otherwise for a further boost in both speed and sensitivity, the fluorescence subtraction can be done in analog, as has been demonstrated for biological samples in Ref. \cite{Barry2016, TurnerDiss2020} and for circuits in Ref. \cite{Webb2022}. With current technologies, measurements here are expected to reach nearly 4 kHz through sub-millisecond exposures. 

\end{itemize}

\begin{figure}[t]
  \centering
  \includegraphics[width=\linewidth]{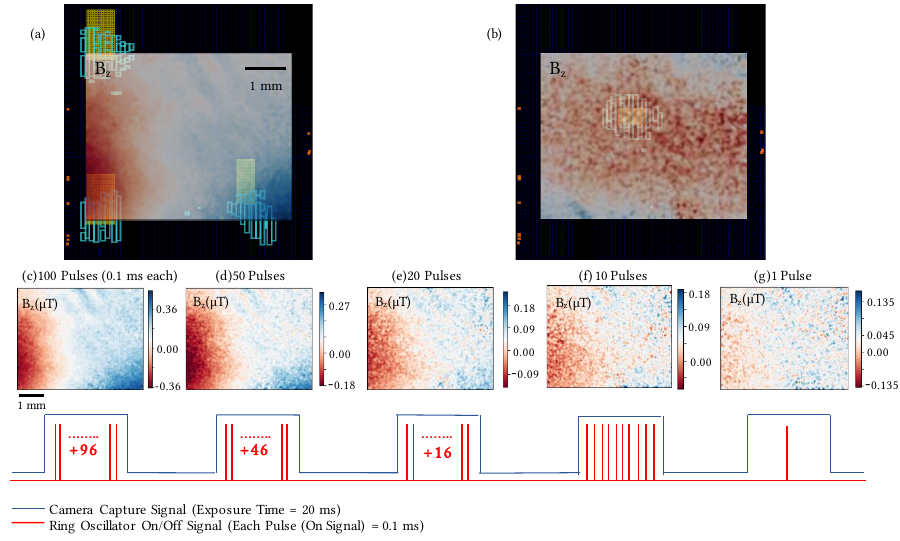}
  \caption{(a) A B$_{z}$ magnetic field image taken of an Artix-7 overlaid on Vivado device layout diagrams showing the location of ring oscillators (ROs) that can vary their activity frequency. For the measurements in (a) and (c)-(g), there are two ROs present on the left side of the field-of-view, and an area of logic dedicated to other background activities, such as debounce, present in the lower right section of the field-of-view. For (a), the ROs are always active.  (b) A B$_{z}$ magnetic field image overlaid on the same Artix-7 FPGA running a single RO, active in the center of the diamond's field-of-view, while the background activity is completely out of the field-of-view. The RO is active half of the time for this image, pulsing 100 times at 0.1 ms per pulse during a window of 20 ms (equal to our exposure time for this set of measurements). (c)-(g) Magnetic field maps of B$_{z}$(µT) with two ROs active on the left side of the field of view as in (a) , scaling down the number of 0.1 ms pulses per 20 ms exposure time from 100 pulses in (c) all the way to a single 0.1 ms pulse in (g), demonstrating the QDM's ability to detect events at a much smaller timescale than the temporal resolution. The scale bar in panel (a) is 1 mm and applies to both panels (a) and (b). The scale bar in panel (c) is 1 mm and applies to panels (c)-(g).}
  \Description{DESCRIBE IMAGES.}
  \label{fig4}
\end{figure}

In order to quantitatively map the magnetic field for the different measurement schemes listed above, the magnetic field measurements are averaged over the entire exposure time for a single measurement. An advantage of this averaging is the ability to detect events at smaller timescales than the temporal resolution of the measurement. In Figure \ref{fig4}, we demonstrated this capability with measurements of an active, thinned (allowing a standoff distance of $\sim$80 µm from the diamond to the die) Artix-7 FPGA with ring oscillators (RO) of varying activity schemes. The sample bitstreams used to program the Artix-7 FPGA were created in Vivado[60].  This tool allowed us to physically constrain regions of ROs to locations on the die that resided within or on the edge of the diamond’s field-of-view (two on the left of the field-of-view in \ref{fig4}(a) and one in the center of the field of view in \ref{fig4}(b)).  Additional control logic was included to allow us to pulse the activity of the ring oscillators with a constant 0.1 ms period and varying frequency (number of pulses per 20 ms exposure period). Figure \ref{fig4}(c)-(g) are magnetic field maps with decreasing numbers of these RO pulses; Figure \ref{fig4}(g) shows the magnetic field map from a single 0.1 ms pulse during a 20 ms exposure period, demonstrating the capability of measuring events at smaller timescales than the temporal resolution. This serves as a proxy for Trojans that may exhibit infrequent activity, demonstrating the QDM's ability to detect this behavior. This experiment also highlights the QDM's ability to detect through some packaging and despite larger standoff distance, as the Artix-7 was only thinned via milling to $\sim$80 µm. Factors such as encapsulation and time-to-result will play a large role in the QDM concept of operations for hardware Trojan detection, as they will affect the potential and cost of detection at various stages of the IC development process. 

\subsection{QDM Concept of Operations}

\begin{figure}[t]
  \centering
  \includegraphics[width=\linewidth]{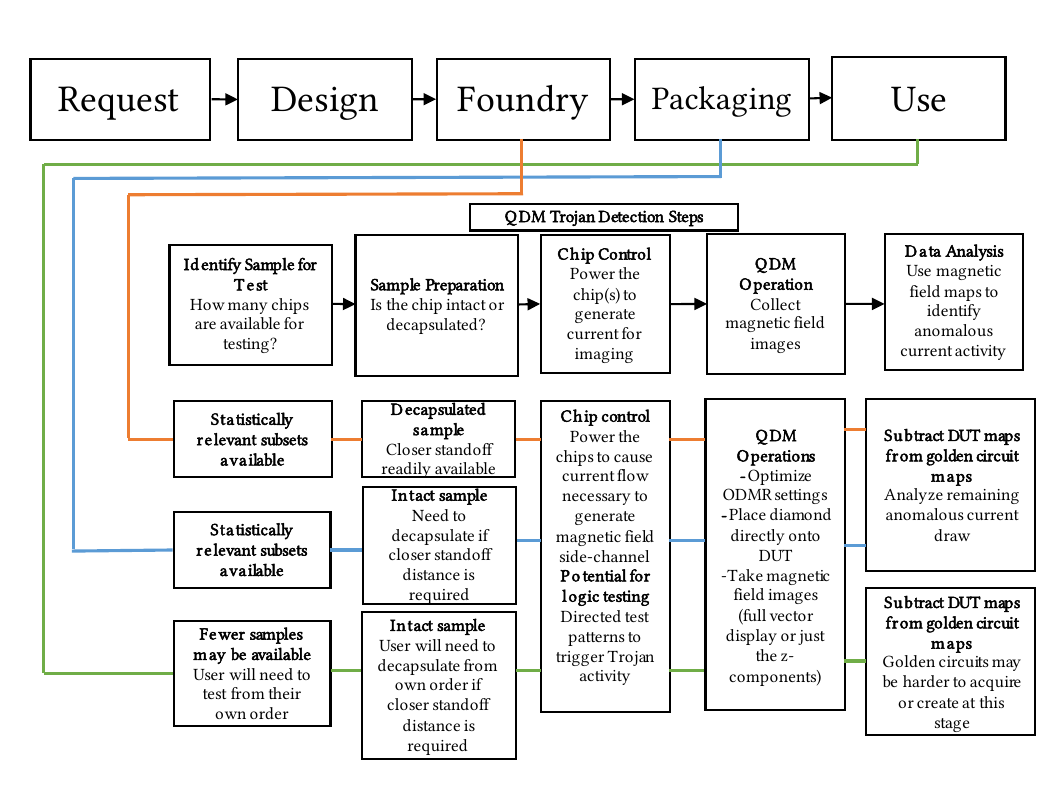}
  \caption{Concept of operations for hardware Trojan detection with the Quantum Diamond Microscope (QDM). The top row is the IC development process flow, with branches from foundry, packaging, and the user where QDM testing could detect compromised components. The orange line from Foundry, blue from Packaging, and green from User indicate the different considerations for each during the QDM hardware Trojan detection steps detailed in the lower three rows. QDM Trojan Detection Steps highlights the main questions and considerations during the detection process, and these are further addressed for foundry, packaging, and user in the three colored paths.}
  \Description{DESCRIBE IMAGES.}
  \label{fig3}
\end{figure}

Figure \ref{fig3} details an envisioned concept of operations (ConOps) for implementing hardware trojan detection with the QDM at different stages of microelectronics component development. When considering QDM use for microelectronics security, it is important to consider the QDM as a unique tool to combine with other techniques (such as SEM and EM probing) to further localize and classify anomalous activity. Moving along the top row of Figure \ref{fig3} from Request to User, Trojan detection becomes more difficult due to sample availability, sample preparation requirements, QDM imaging requirements, data analysis considerations, as well as golden circuit availability. When identifying samples and preparing them for testing as depicted in the first and second columns of the lower four rows, the IC development stage determines how many samples are available and the difficulty expected for detection via magnetic field imaging (i.e. whether samples are encapsulated). 

As listed in the second column of Figure \ref{fig3} under QDM Detection Steps, QDM spatial resolution will depend on standoff distance of the diamond sensor from the current sources. If testing immediately after fabrication and before packaging, stand-off distances can be minimized resulting in better spatial resolution, which is important here because Trojan logic can be small in comparison to the intended logic. If more information on the type and effect of tampering is needed, localization with magnetic field imaging can help to know where to look with other techniques like scanning electron microscopy (SEM). Once packaged, chips will either need to be decapsulated or testing parameters will need to be adjusted for proper through-package imaging. 

As shown in the third column of Figure \ref{fig3}, chip control and power are necessary to induce current flow in the DUT, which is the source of the magnetic field to be measured with the QDM, and implementation depends heavily upon chip specifications, function, and application. The potential to aid Trojan detection with the QDM through logic testing exists since the Trojan's payload can be intentionally triggered or activated, potentially increasing the likelihood of detection \cite{ChakrabortyMERO2009,Banga2009}. If there is a known Trojan suspected for a specific chip function, one could attempt activation with a set of directed test patterns before conducting side-channel analysis with the QDM. This could greatly increase detection efficiency for Trojans in which the payload logic is much larger than the trigger logic. 

Having completed the preparatory steps, the device can now be measured via QDM operation as shown in column 4 of Figure \ref{fig3}. Running the instrument and conducting data analysis can be carried out by technical staff that are capable of running other characterization techniques. Measurement parameters for the QDM depend on each of these earlier steps, as operators must consider the number of samples, the packaging or lack thereof, and the chip control protocol. Other considerations for QDM operation include the collection time for each measurement, which depends on the desired time-to-result and is multiplied by the number of samples to be measured, and whether or not comparisons to the golden chip include all vector dimensions or just the magnetic field in Z. 

Finally, as highlighted by the last column of Figure \ref{fig3}, design of any data analysis protocol must consider each step in this process in order to properly analyze the magnetic field maps resulting from QDM operation. This will usually include the averaging of multiple measurements to improve signal to noise ratio and the subtraction of the golden circuit magnetic field maps from the magnetic field maps of the DUT to isolate anomalous activity. Here the effects of process variation will have to be accounted for and the device physics will need to be well understood to distinguish anomalous activity from simple process variation. The device specifications will play a key role in this determination, as smaller process sizes can lead to large increases in process variation.

\section{Hardware Trojan Effects on Current Side Channels}

In this section we conduct a rough estimation of the effects of a hardware Trojan on the current (magnetic) side channel to determine the capabilities of the QDM for Trojan detection on varying types of hardware. Every hardware Trojan introduces a deviation from the original design's circuitry. Since digital logic and memory are composed of transistor circuits, any change to a design's functionality results in a change to the current distribution. Each of these changes contribute to the current distribution which gives rise to the magnetic side channel measured by the QDM. Estimating the effect of these contributions allows for a projection of QDM performance at varying IC process sizes and QDM total measurement times. 

The amount of current activity introduced by each transistor is dependent on the level of IC technology (i.e. process size). This estimation is carried out with $I=CV \alpha f$ where $I$ is the current, $C$ is the gate capacitance, $V$ is the voltage, $\alpha$ is the activity factor, and $f$ is the switching frequency. Activity factor is the probability of a gate switching its output from 0 to 1, which is heavily dependent on the intended activity of the gate. For each hardware Trojan and chip implementation of concern, in cases where spatial resolution of the current is desired, the smallest possible change in an area should be considered for determining detection capability. This is due to the fact that Trojan power consumption can be spread over large areas of the circuit, making changes to a spatially resolved image non-discrete. Trojan size affects power consumption, and can be viewed as a combination of the physical alterations to the circuitry and the propagation of operational changes throughout the circuitry due to the payload. When spatially resolving current effects, larger Trojan size can be treated as presenting more opportunities for detection of the Trojan with the QDM. The sensitivity of a given detection protocol can thus be assessed through consideration of the smallest possible changes to current activity due to the Trojan's presence. An analysis of current change due to single transistors can demonstrate sensitivity needs for Trojan detection on devices of varying process size. 

\begin{figure}[t]
  	\centering
	\includegraphics[width=\linewidth]{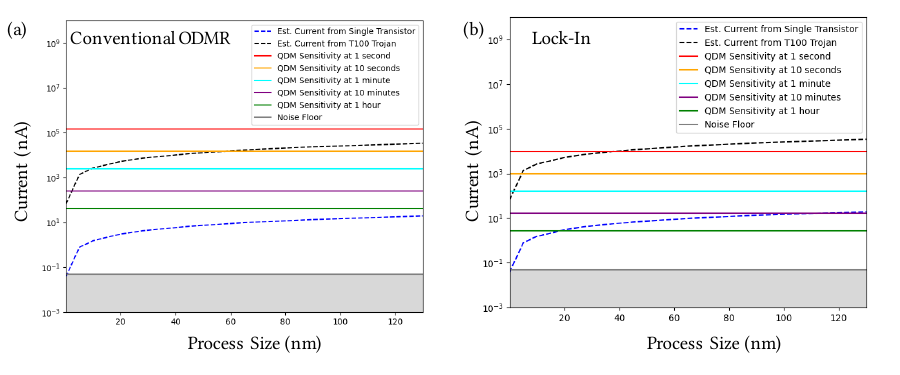}
	\caption{Integrated Circuit (IC) Process Size (nm) vs. Current (nA). The dashed blue fit line indicates the current introduced by a single transistor at each process size, and the dashed black fit line indicates the total current introduced by an AES-T100 Trojan benchmark. These current estimations can then be compared to the QDM sensitivities at various temporal modes. (a) shows QDM sensitivities with standard setup (sweeping many microwave frequencies) at a 10 µm stand-off distance and (b) describes an optimized setup making use of digital lock-in at a 10 µm stand-off distance. The environmental noise floor is indicated by the greyed out area, preventing detection in this region without electromagnetic shielding.}
	\Description{DESCRIBE IMAGES.}
 	\label{plot3}
\end{figure}

Figure \ref{plot3} shows the estimated current introduced per transistor (dashed blue fit line) and the estimated total current introduced for the AES-T100 Trojan benchmark from TrustHub (dashed black fit line). These trust benchmarks, each defined with a set Trojan location and size, were developed to allow for the comparison of hardware Trojan detection methods. The AES-T100 benchmark describes an always-on hardware Trojan with a side-channel leakage payload \cite{Salmani2013, Shakya2017}. It leaks a key from a chip running an AES cryptographic algorithm without needing to be triggered, utilizing a set number of logical elements. Other TrustHub benchmarks utilize similar numbers of logical elements for their trigger monitoring alone \cite{Ashok2022}. The dashed black lines in Figure \ref{plot3} indicate the total current difference due to the AES-T100 when measured from the chip as a whole, which could be accomplished by the QDM if the signals from each pixel were read out cumulatively although this would sacrifice the QDM's ability to leverage spatial resolution of chip features. In practice, when spatially resolving the chip activity the current introduced by a useful Trojan would most likely fall between these two estimations. 

These estimations allow for rough calculations of the amount of time needed to measure to ensure detection of the relevant current activity for each process size. It should also be noted that the 10 µm standoff distance assumed for the compared QDM sensitivities would typically only be true for a decapsulated IC; Figure \ref{plot1} shows the relevant sensitivities at farther standoff distances. Figure \ref{plot3} highlights the fact that the QDM's sensitivity at moderate measurement times is adequate to measure the possibly small changes in current due to hardware Trojans on a wide range of process sizes. Further experimentation with the QDM could both confirm the relevant levels of current activity for hardware Trojan benchmarks as well as demonstrate detection capabilities at these timescales.

\section{Conclusion}

In this work, we sought to identify the capabilities and limits of the QDM for detection of hardware Trojans. We considered a threat model of a compromised foundry that produced a subset of hardware Trojan implanted chips that could be detected at any step of the IC development process after fabrication. These stages include the foundry and packaging steps, as well as detection attempts by a user before implementation of a chip. We highlighted the categories of hardware Trojans and methods for detection, including logic testing and side-channel analysis. The QDM's side-channel analysis capability shows promise for detection of hardware Trojans at any stage of the development process after fabrication. 

We collected and discussed relevant QDM metrics, including sensitivity, spatial resolution, time-to-result and field-of-view, for hardware Trojan detection. The QDM spatial resolution of $\sim$ 10 µm demonstrated in Ref. \cite{Turner2020} is sufficient for activity localization on 28 nm process size chips. We made sensitivity projections for varying measurement times based on conventional and lock-in QDM measurement schemes. We demonstrated rare (0.1 ms per 20 ms exposure period) event detection on a thinned Artix-7, which can serve as a proxy for similar Trojan behavior. The projected sensitivities at various measurement times were then compared to estimated current activity due to a single transistor and an AES-T100 Trojan benchmark. This estimation showed the QDM to be sensitive enough at moderate measurement times to detect Trojan activity on chips of various process sizes. 

Hardware Trojan detection remains an elusive problem due to the wide variation possible when implementing potential Trojans. QDM detection of hardware Trojan activity appears a valid tactic to add to the toolbox of embedded security, showing promise for detecting and localizing anomalous activity on chips of varying process sizes and at varying stages of development. Further experimentation needs to be done to establish the capability of the QDM to detect hardware Trojans through-package, as well as confirm the total measurement time needed for detection at varying IC process sizes. 

\section{Acknowledgements}

The Homeland Security Act of 2002 (Section 305 of PL 107-296, as codified in 6 U.S.C. 185), herein referred to as the “Act,” authorizes the Secretary of the Department of Homeland Security (DHS), acting through the Under Secretary for Science and Technology, to establish one or more federally funded research and development centers (FFRDCs) to provide independent analysis of homeland security issues. MITRE Corp. operates the Homeland Security Systems Engineering and Development Institute (HSSEDI) as an FFRDC for DHS under contract 70RSAT20D00000001. 

The HSSEDI FFRDC provides the government with the necessary systems engineering and development expertise to conduct complex acquisition planning and development; concept exploration, experimentation and evaluation; information technology, communications and cyber security processes, standards, methodologies and protocols; systems architecture and integration; quality and performance review, best practices and performance measures and metrics; and, independent test and evaluation activities. The HSSEDI FFRDC also works with and supports other federal, state, local, tribal, public and private sector organizations that make up the homeland security enterprise. The HSSEDI FFRDC’s research is undertaken by mutual consent with DHS and is organized as a set of discrete tasks. This report presents the results of research and analysis conducted under: 70RSAT22FR0000021, “DHS Science and Technology Directorate TCD QIS Capabilities.” 

The results presented in this report do not necessarily reflect official DHS opinion or policy.

Approved for Public Release; Distribution Unlimited. MITRE Public Release Case Number 23-4322. 

\bibliographystyle{ACM-Reference-Format}
\bibliography{qdm-review-draft}

\appendix

\section{Appendix: NV Physics}

NV centers come in three possible charge states: $NV^{-}$, $NV^{+}$, and $NV^{0}$. The charge state used for imaging integrated circuits with the QDM is $NV^{-}$. Each NV center displays a spin-1 triplet electronic ground state and must lie in one of four possible orientations in the crystallographic diamond structure. The spin states are identified as $m_{s} = 0$ and $m_{s} = \pm 1$. The $m_{s} = \pm 1$ states are degenerate in the absence of a magnetic field, and experience Zeeman splitting in the presence of a magnetic field. The four possible orientations each measure the magnetic field changes in a different direction, providing the basis for vector magnetometry with the NV centers. A precisely aligned bias magnetic field can be applied to split the $m_{s} = \pm 1$ states of all four orientations to conduct precise and sensitive magnetometry. 

\begin{figure}[t]
  	\centering
	\includegraphics{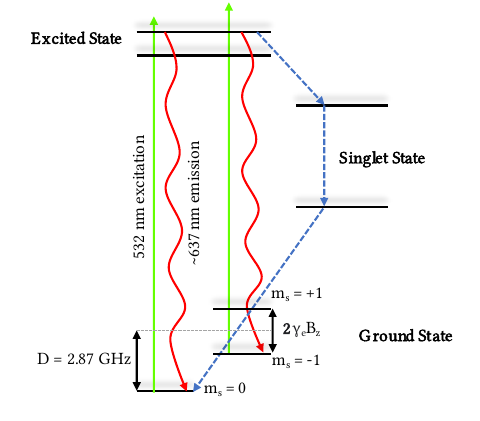}
	\caption{Level structure of a Nitrogen vacancy center in diamond. Incident 537 nm light causes excitation of both $m_{s} = 0$ and $m_{s} = \pm 1$ from ground to the excited state. Once excited, NV centers in the $m_{s} = 0$ state transition back to ground, emitting $\sim637$ nm light. NV centers in the $m_{s} = \pm 1$ state are more likely to drop into a singlet state, after which they decay to the $m_{s} = 0$ state. In the absence of a magnetic field, the zero-field splitting parameter D = 2.87 GHz is the resonant frequency to go from $m_{s} = 0$ to $m_{s} = \pm 1$. In the presence of a magnetic field, $m_{s} = \pm 1$ resonances split. Ignoring nuclear spin and in the presence of a magnetic field constrained to the z-axis, this splitting is $2 \gamma^{}_{e}B_{z}$.}
	\Description{ADD DESCRIPTION.}
 	\label{levels}
\end{figure}

Figure \ref{levels} details the effective level structure of an $NV^{-}$ center. Incident green ($\sim$532 nm) laser light  excites NV centers from their ground state to an excited spin-triplet state. Excited NV centers in the $m_{s} = 0$ and $m_{s} = \pm 1$ states that decay to the ground state fluoresce in the 600-850 nm band \cite{Barry2020}. From the excited state, NV centers in the $m_{s} = \pm 1$ state are more likely to decay into a singlet state. From this singlet-state they do not fluoresce in the 600-850 nm band. The $\sim$140-200 ns lifetime of this singlet state is much longer than the 13 ns lifetime of the excited state. Thus, by detecting 600-850 nm light emitted by the diamond, we can determine the proportion of NV centers in the $m_{s} = 0, \pm 1$ states. 

The distribution of $m_{s} = 0, \pm 1$ states is also determined by incident resonant microwaves on the diamond. The zero-field-splitting parameter $D \approx 2.87$ GHz splits $m_{s} = 0$ and $m_{s} = \pm 1$ as the resonant frequency when no magnetic field is present. Zeeman splitting of the $m_{s} = \pm 1$ states caused by magnetic fields can be read out by determining the shifting of resonant frequency. This is the basis of Optically Detected Magnetic Resonance (ODMR). For continuous wave ODMR the 532 nm laser runs continuously, and a set of microwave frequencies are swept through. Dips in fluorescence are detected by a photodiode or camera, and the frequency separation between these dips indicates magnetic field magnitude.

\section{Appendix: Signal-to-Noise Ratio}

\begin{figure}[H]
  	\centering
	\includegraphics[width=\linewidth]{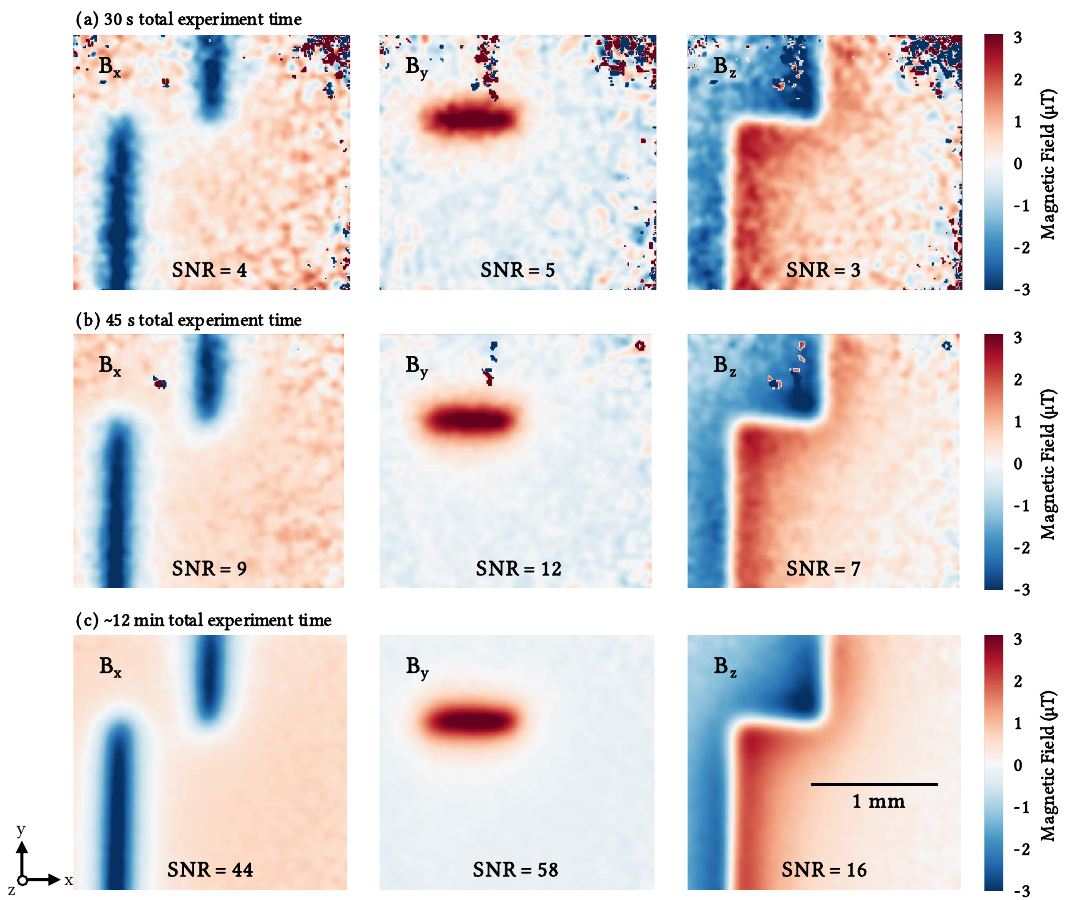}
	\caption{Quantum Diamond Microscope (QDM) magnetic field measurements taken on a custom printed circuit board (PCB) that was developed as a testbed to explore various aspects of the instrument’s magnetic imaging capabilities. Here, the QDM’s diamond sensor is placed on the top layer of the PCB where it is separated from a current trace embedded 18 µm deep in the sample. The trace is biased with 1 V with a 1000 $\Omega$ resistor so that 1 mA flows through the wire and the resultant magnetic field’s vector components (B$_{x}$, B$_{y}$, and B$_{z}$) are imaged with the QDM. Various experimental parameters can be controlled, which affect measurement time in different ways. Measurements in panel (a) took 30 seconds, measurements in panel (b) took 45 seconds, and measurements in panel (c) took $\sim$12 minutes, resulting in images with various signal-to-noise ratios (SNR). While relatively high current is flowing in this sample, the images suggest that useful information for localizing activity can be extracted from QDM magnetic field images with an SNR somewhere in the 5-10 range, allowing for significantly reduced experiment wall time. Extended measurements in time only act to drive down measurement noise. For lower SNR images, blotchy artifacts are common and result from sub-optimal curve fitting to noisy optically detected magnetic resonance spectra (example spectrum shown in Figure \ref{fig2}(d)). The scale bar in panel (c) is 1 mm and applies to all panels.}
	\Description{ADD DESCRIPTION.}
 	\label{snrmaps}
\end{figure}

\end{document}